\documentclass[twocolumn,showpacs,prl]{revtex4}
\usepackage[dvipdfm]{graphicx}
\usepackage{amsmath}
\usepackage{amssymb}
\usepackage{amsfonts}

\begin{document}

\title{Hopping transport of interacting carriers in
disordered organic materials}

\author{Sergey V. Novikov}

\affiliation{A.N. Frumkin Institute of Physical Chemistry and
Electrochemistry, Moscow 119991, Russia}


\begin{abstract}Computer simulation of the hopping charge transport
in disordered organic materials has been carried out explicitly
taking into account charge-charge interactions. This approach
provides a possibility to take into account dynamic correlations
that are neglected by more traditional approaches like mean field
theory. It was found that the effect of interaction is no less
significant than the usually considered effect of filling of deep
states by non-interacting carriers. It was found too that carrier
mobility generally increases with the increase of carrier density,
but the effect of interaction is opposite for two models of
disordered organic materials: for the non-correlated random
distribution of energies with Gaussian DOS
 mobility decreases with the increase of the interaction strength, while for the
model with long range correlated disorder mobility increases with
the increase of interaction strength.
\end{abstract}

\maketitle                   

\section{Introduction}

Hopping transport in disordered organic materials has been
extensively studied for the case of low density of carriers, but
our understanding of charge transport for the case of high carrier
density is not adequate; theoretical studies of the effect of
carrier density are scarce \cite{density1,density2}. High density
of carriers could affect drift mobility $\mu$ in opposite ways.
Small fraction of carriers could occupy deep states, thus
providing a possibility for remaining carriers to avoid trapping
and acquire much higher mobility. At the same time, charge-charge
interactions could provide an additional energetic disorder in the
material. This is indeed the case for the simplest model where all
charges except one are immovable \cite{PF_static}. In this case,
the greater is the density of static charges (i.e. energetic
disorder), the smaller is the mobility.


Usually, in theoretical studies the mean field approximation has
been used and charge-charge interaction has been totally neglected
\cite{density1,density2}. This means that these studies dealt only
with the effect of filling of deep states: it is assumed that the
effects of interaction could be later effectively included via the
mobility dependence on the mean local electric field
$\left<\vec{E}_{\rm loc}\right>$, which in turn is connected to
the mean local charge density $\rho$ by the Poisson equation
\begin{equation}\label{Poisson}
{\rm{div}} \left<\vec{E}_{\rm loc}\right> =
\frac{4\pi}{\varepsilon} \rho,
\end{equation}
where $\varepsilon$ is a dielectric constant of the material. This
line of reasoning totally neglects dynamic correlations. In
addition, quite frequently true quasi-equilibrium mobility is
formed by the averaging over large domains of the disordered
material (see, for example, Ref. \cite{density2}). In this
situation the very conception of a local (but uniform in space)
mobility  is invalid. For this reason in this paper we describe a
direct dynamic simulation of hopping monopolar transport of
\textit{interacting} carriers.

\section{Model}
\subsection{General description}
For the model of interacting carriers the total site energy of a
carrier is a sum of a static random contribution due to the static
intrinsic energetic disorder and fluctuating contribution due to
the interaction with all other hopping charges. Tremendous
difficulty of
 the simulation is
a necessity to recalculate site energies and hopping probabilities
after every hop. In addition, if we consider moderate or high
density of charge carriers, then
 they affect the distribution of average electric field,
which in turn leads to the non-uniform average charge density. In
this situation we cannot use in simulations a finite basic sample
 with periodic boundary conditions. The length of the basic sample
 across the device must be equal to
the thickness of the transport layer. This inevitably means that
we have to include charge injection in our simulation and analysis
of the simulation data becomes much more complicated.

For these reasons we undertook simulation for the simplified model
of a transport layer: we assumed that a static compensating charge
of the opposite sign is uniform\-ly distributed in the transport
layer, with the density being equal to the average density of
carriers. In this model the mean local electric field and carrier
density are uniform in space and
 $\left<\vec{E}_{\rm loc}\right>$ is equal to the applied field $\vec{E}$.
 Hence, simulation for the moderate
basic sample with periodic boundary conditions could be carried
out. This particular model is a variant of the "jelly model", very
popular for a study of effects of charge-charge interactions in
metals.

In order to minimize the necessary runtime even more we
$\hskip2pt$ performed $\hskip2pt$ simulations $\hskip2pt$ for
$\hskip2pt$ rather $\hskip2pt$ high temperature
 $kT/\sigma=0.3$ (here $\sigma^2$ is a variance of
static intrinsic disorder; typically, in organics $\sigma\simeq
0.05 - 0.1$ eV \cite{BW}) and took into account only hops to
nearest neighbors of every site (we considered a simple cubic
lattice). Site energies and hopping probabilities have been
recalculated only for nearest neighbors of sites occupied with
carriers. This assumption seems to be very good approximation for
disordered organics where $\alpha a \simeq 5 - 10$ \cite{GDM1},
here $a \simeq 1$ nm is a distance between nearest neighbors
(lattice scale in a lattice model of a material) and $\alpha$ is
an inverse localization radius of a wave function of a transport
site. Miller-Abrahams (MA) hopping rate \cite{MA_rate} has been
used for all simulations. In the calculation of electrostatic site
energies a well-established Ewald approach has been used
\cite{ewald}.

Simulations for two models of a disordered organic material have
been carried out: the Gaussian Disorder Mo\-del (GDM) where the
static intrinsic energetic disorder in the material is represented
by a spatially uncorrelated distribution of random energies with
the Gaussian DOS \cite{GDM1} and the model of dipolar glass (DG)
where site energies are strongly correlated \cite{DG1}. The later
model is better suited to describe organic materials because
contributions of static electrostatic sources (of dipolar or
quadrupolar nature) to site energies are inevitably strongly
spatially correlated due to the long range nature of electrostatic
interaction in organic materials \cite{clusters,QP}.

Majority of simulations have been performed for the basic sample
of a simple cubic lattice with the size $L=40a$. Several checks
have been made to test an accuracy of the simulation by using the
basic sample with size $L=80a$. $\hskip2pt$ Even $\hskip2pt$ for
$\hskip2pt$ the DG model $\hskip2pt$ and $\hskip2pt$ rather
$\hskip2pt$ weak $\hskip2pt$ field
 $eaE/\sigma\approx 0.1$ (these are the situations
where mobility is very sensitive to the size of a sample) a good
agreement has been found with the case $L=40a$.

Simulation for a particular set of relevant parameters was started
with some arbitrary distribution of carriers and carried out until
a stationary state with constant mean carrier velocity has been
reached. We have checked that the particular initial distribution
of carriers affects only details of the relaxation process but not
the value of the asymptotic velocity.

\subsection{Waiting time simulation}
General features of the simulation procedure are very similar to
those described in Refs. \cite{GDM1,DG1}, apart from the
modification of the waiting time calculation. For the only carrier
a waiting time $\tau$ before the next hop is calculated as
\begin{equation}\label{waiting time}
\tau=-\frac{1}{\Gamma}\log\gamma,
\end{equation}
where $\gamma$ is a random number with the uniform distribution in
[0,1] and $\Gamma$ is the total hopping rate for the site where
carrier is located at the moment \cite{GDM1}. For many interacting
carriers this procedure has to be modified because for any
particular carrier, waiting for a hop, the total hopping rate now
depends on time, reflecting hopping of other carriers.

A necessary modification
 could be introduced in the following way. At the start, independent
 $\gamma_n$ have been generated for all carriers, and the set of $\tau_n$
 has been calculated using Eq.
(\ref{waiting time}), as well as the set of initial residuals
$R_n=-\log\gamma_n$. Then the carrier with smallest waiting time
$\tau_{\rm min}$ made a hop, its new position has been found by
the usual way \cite{GDM1}, and the time counter was advanced by
$\tau_{\rm min}$. For this carrier a new random number
$\gamma^{\rm new}$ was generated. For all other carriers new
residuals
\begin{equation}\label{residuals}
R_n^{\rm new}=R_n -\Gamma_n\tau_{\rm min}
\end{equation}
have been calculated and stored (for the hopped carrier a new
residual is just $R^{\rm new}=-\log \gamma^{\rm new}$).  Then all
site energies and hopping rates have been recalculated, the new
set of waiting times have been calculated
\begin{equation}\label{waiting time_2}
\tau_n^{\rm new}=\frac{R_n^{\rm new}}{\Gamma_n^{\rm new}},
\end{equation}
and the whole procedure has been repeated.

\section{Dangers of mean field approximation}
\begin{figure}[floatfix]
\begin{center}
\includegraphics[width=3in]{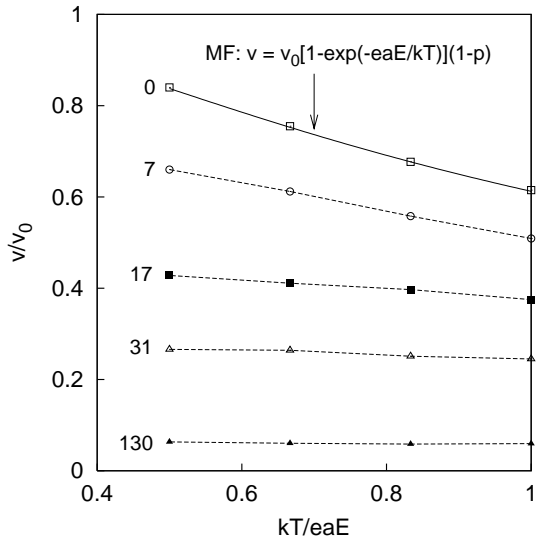}
\end{center}
\caption{Temperature dependence of the mean carrier velocity $v$
for the MA hopping rate in the case of no intrinsic disorder
(dynamic simulation, points). Upper solid line is the mean field
result. Numbers near the curves show effective charge-charge
interaction parameter $e/\varepsilon a^2 E$. Average occupation
fraction is $p=0.047$.} \label{Charges}
\end{figure}

To illustrate possible dangers of the mean field approximation we
provide results of the transport simulation for the simplest case
of no intrinsic static disorder (carriers, hopping in the empty
lattice). In this case the mean field solution for the
 carrier occupation fraction at every site is just a constant equal
 to the average occupation fraction $p$. For the MA hopping rate
an average carrier velocity for the case of simple cubic lattice
and hopping to nearest neighbors only  is
\begin{equation}\label{MF_no_disorder}
v_{\rm MF}=v_0\left[1-\exp\left(-eaE/kT\right)\right](1-p),
\hskip10pt v_0=a\Gamma_0,
\end{equation}
where two terms in brackets represent contributions of forward and
backward hops, factor $1-p$ takes into account occupation of the
final site, and  $\Gamma_0$ is a hopping frequency. Remarkable
property of Eq. (\ref{MF_no_disorder}) is its independence on the
interaction strength. Indeed, in our model in the absence of
static intrinsic disorder an average total charge density is
exactly zero everywhere, thus all interactions enter into
consideration only because of dynamic correlations. Results of the
simulation are shown in Fig. \ref{Charges}. For strong interaction
deviation with Eq. (\ref{MF_no_disorder}) is significant. Note
also, that for $e/\varepsilon a^2 E=0$ agreement between Eq.
(\ref{MF_no_disorder}) and simulation data is excellent due to the
small average occupation fraction $p \ll 1$ used in simulation;
here dynamic correlations are not important.

\begin{figure}[floatfix]
\begin{center}
\includegraphics[width=3in]{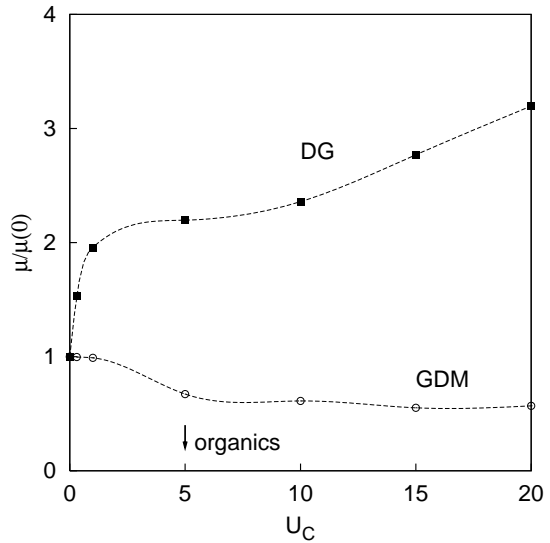}
\end{center}
\caption{ Dependence of the carrier mobility on
the effective strength of the charge-charge interaction
$U_C=e^2/\varepsilon a \sigma$ for $L=40a$, $p=0.008$,
$kT/\sigma=0.3$, and $eaE/\sigma=0.1$. Black squares show the data
for the DG model and empty circles show results for the GDM. For a
typical disordered organic material $U_C \simeq 5$. Here $\mu(0)$
is the mobility for $U_C=0$.} \label{Int_effect}
\end{figure}

\section{Results of dynamic simulation}
\subsection{General effect of interaction}

We found that in our model for not very high occupation fractions
$p \leq 0.1$ carrier drift mobility increases with $p$, exactly as
in the case of non-interacting carriers \cite{density1,density2}.
Yet modification provided by the interaction is still significant
(see Fig. \ref{Int_effect}). There is a striking difference
between the effects of interaction (i.e., carrier-carrier
repulsion) on the mobility in the DG model and GDM. In the DG
model repulsion between carriers makes mobility even greater than
in the case of no interaction, while for the GDM the opposite
situation takes place. This difference is not surprising. Indeed,
in the DG charge transport is significantly affected by carrier
trapping in deep and broad valleys of the energy landscape (good
qualitative description is provided in Ref. \cite{DG_picture}). If
a carrier is trapped by a valley, then the whole valley with many
sites having low energies becomes blocked for other carriers
because of repulsion. Thus, filling of the deep states is much
more effective in correlated landscape if carrier repulsion is
taken into account. This is the reason for the increase of carrier
mobility in DG. No such effect takes place for the GDM, and here,
evidently, the effect of charge-induced energetic disorder is
responsible for the decrease of mobility in comparison to
non-interacting case.

\begin{figure}[floatfix]
\begin{center}
\includegraphics[width=3in]{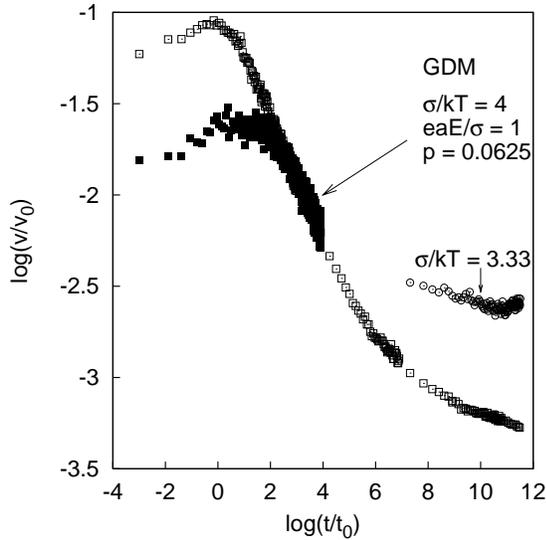}
\end{center}
\caption{ Relaxation of the mean carrier velocity in the GDM for
$\sigma/kT=4$, $eaE/\sigma=1$, and $p=0.0625$ (squares). Empty
squares show relaxation for random initial locations, while filled
squares show relaxation for the case, when initial locations have
been taken at the minimal energy positions. Circles shoe the
realaxation for higher temperature $\sigma/kT=3.33$.}
\label{relaxation}
\end{figure}

This particular result disagrees with the result of a recent paper
by Zhou \textit{et al} \cite{Zhou}. They found that in the GDM
carrier interaction \textit{enhances} mobility in comparison to
the case of no interaction if $\sigma/kT \gg 1$. This is opposite
to our findings. Quite probably, the disagreement stems from the
under-relaxation of the initial (random) carrier configuration
used in Ref. \cite{Zhou}; the relaxation process is pretty slow
for interacting carriers if $\sigma/kT \gg 1$. We cannot make more
detailed comparison because typical relaxation times are not
provided in Ref. \cite{Zhou} (in fact, even the strength of
carrier repulsion is not provided). Our data indicates, for
example, that for $\sigma/kT=4$ relaxation is not completely over
even for $t/t_0=1\times 10^5$ (see Fig. \ref{relaxation}); at that
time carrier has already travelled in the field direction the
distance of $\simeq4\times 10^3 a$.

Remarkable feature of Fig. \ref{relaxation} is a universality of
the late relaxation stage. Very early relaxation is different for
the initial random distribution and minimal energy distribution
(where every carrier was placed at the site where the total
energy, provided by the intrinsic disorder and all previously
added carriers, has a minimum), but after $t/t_0\simeq 10$
relaxation curves merge into a single curve.

\begin{figure}[floatfix]
\begin{center}
\includegraphics[width=3in]{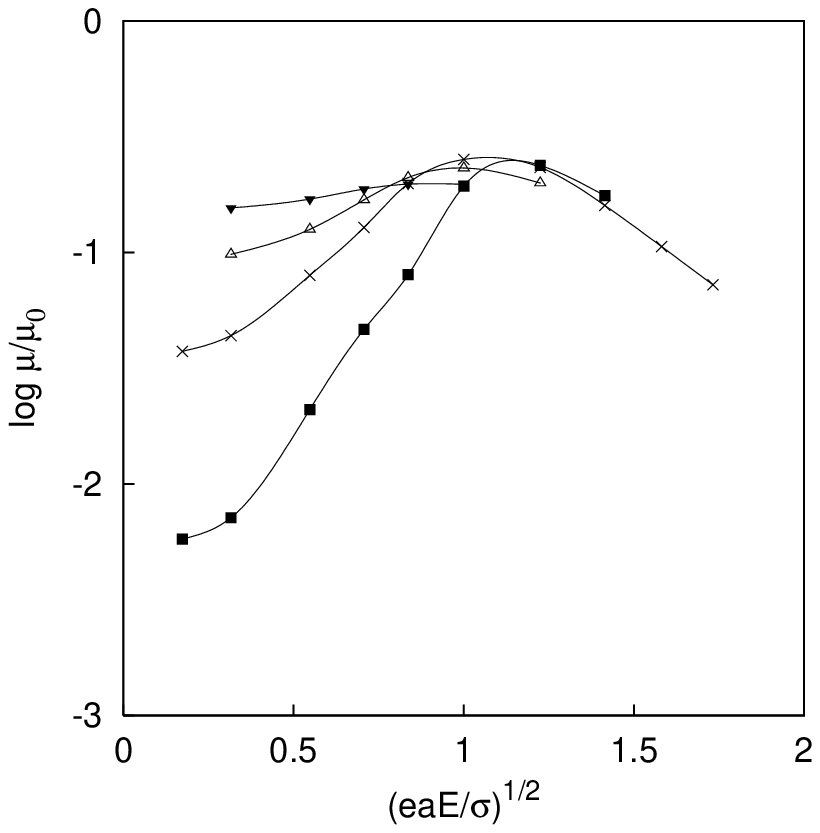}
\end{center} \centering a
\begin{center}
\includegraphics[width=3in]{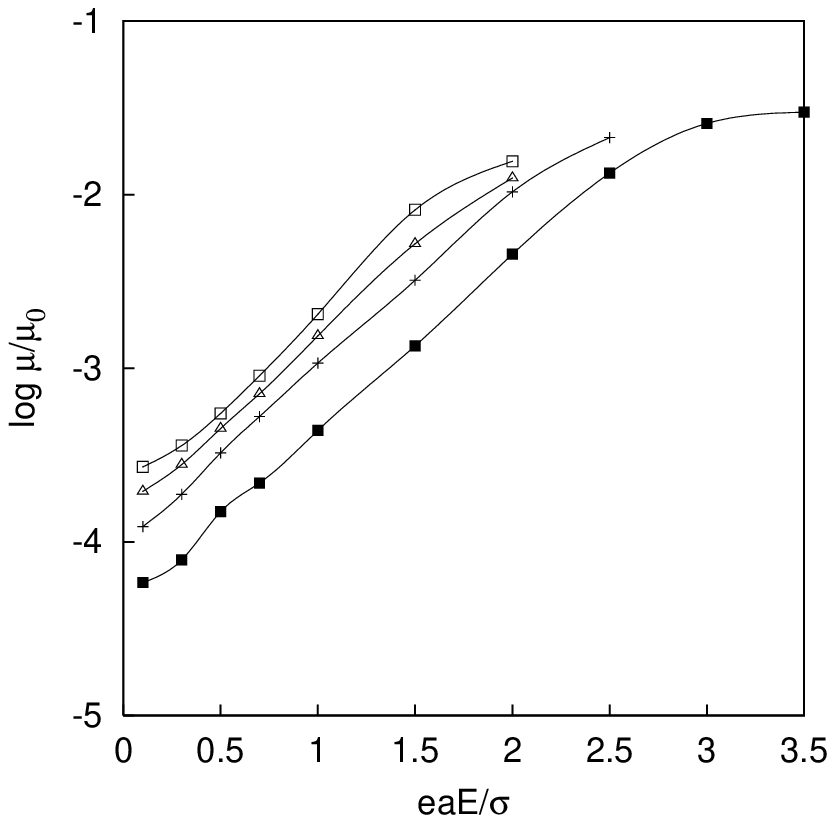}
\end{center}
 \centering b
\caption{ Mobility field dependence for
 $U_C=5$, $kT/\sigma=0.3$, and different values of $p$: 0.0016, 0.008, 0.016, and 0.032,
 from the bottom curve upward, correspondingly. Figure (a) shows
 simulation data for the DG model, and the figure (b) shows data for the GDM, correspondingly.
 Here $\mu_0=ea^2\Gamma_0/\sigma$. If $a= 1$ nm and $\sigma= 0.1$ eV, then
 $eaE\approx 1$ for $E=1\times 10^6$ V/cm.}
\label{mu(E)}
\end{figure}

\subsection{Mobility field dependence}
It is very well known that the GDM and DG model provide very
different dependences of the mobility on applied electric field in
the case of low carrier density. In the GDM $\log\mu\propto E$ and
in DG $\log \mu \propto \sqrt{E}$ \cite{DG1}. These dependences
remains valid for moderate $p$ too (see Fig. \ref{mu(E)}).

Significant difference between models could be found in the
 dependence of the slope of $\mu(E)$ on $p$. General
tendency for the DG model is that transformation of the mobility
curve with the increase of $p$ roughly resembles the corresponding
transformation of the curve with the increase of $T$ (see, for
example, Fig. 1 in Ref. \cite{prl}): with the increase of $p$
mobility becomes greater and the slope of the mobility field
dependence becomes smaller. This is not the case for the GDM: here
only mobility curve moves upward but the slope remains
approximately constant.

This difference could be easily understood. Field dependence of
$\mu$ in the GDM is governed by the carrier escape from deep
states to the nearest transport sites usually having much higher
energy and could be described by the shift $\Delta U = eER$ of
positions of transport levels in applied field $E$
\begin{equation}\label{mu(E)_GDM}
\log \mu \propto \frac{eER}{kT},
\end{equation}
where $R\simeq a$ is a typical distance between transport sites.
Random charge distribution provides a smooth random energy
landscape superimposed on the intrinsic disorder, but typical
additional variation of energy at the scale $a$ is negligible for
small $p$. Hence, estimation (\ref{mu(E)_GDM}) remains valid and
the slope of the mobility curve does not depend on $p$.

Situation in the DG model is quite different. Here mobility field
dependence is governed by the carrier escape from wide valleys,
and the size of critical valleys, capable of keeping a carrier for
the longest time, depends on applied field \cite{DG_picture}
\begin{equation}\label{Rcrit}
R_{\rm cr}\simeq \left(a\sigma^2/ekTE\right)^{1/2}.
\end{equation}
Estimation (\ref{mu(E)_GDM}) should be replaced by
\begin{equation}\label{mu(E)_DG}
\log \mu \propto \frac{eER_{\rm cr}}{kT}\propto \sigma
\left(\frac{eaE}{(kT)^3}\right)^{1/2},
\end{equation}
and this result explains the origin of the famous Poole-Frenkel
mobility field dependence observed in organic materials for the
case of low density of carriers. If we increase the density of
carriers, then at first they fill these critical traps, because
the release time is maximal here. Hence, transport of more mobile
carriers is governed by valleys (clusters) with the size that
differs from $R_{\rm cr}$. We can safely assume that the relevant
traps has $R < R_{\rm cr}$, because the distribution of clusters
on size $R$ decays with $R$. Hence, effective size $R$ in Eq.
(\ref{mu(E)_DG}) is smaller than $R_{\rm cr}$, and, naturally, the
slope of the mobility field dependence should be smaller than the
slope of the corresponding  curve in the case of $p\rightarrow 0$
 (and should decrease with the increase of
$p$), exactly as it was observed in simulation.

\begin{figure}[floatfix]
\begin{center}
\includegraphics[width=3in]{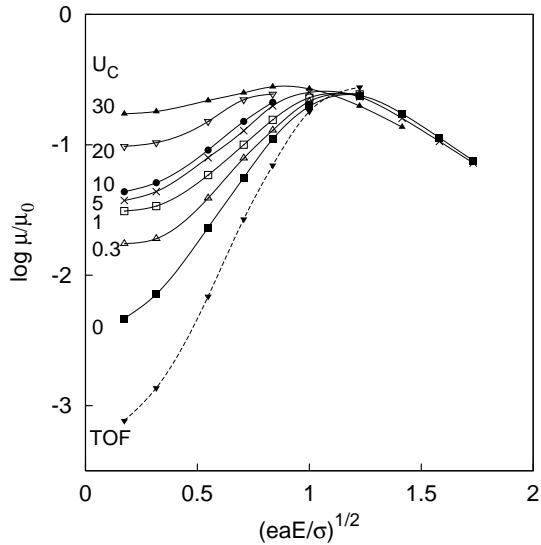}
\end{center}
\caption{ Mobility field dependence for the DG
model for
 different values of $U_C$, indicated at the curves. Other parameters are:
 $kT/\sigma=0.3$ and
  $p=0.008$. The very bottom curve corresponds to the data of a conceivable
  TOF experiment with $p\rightarrow 0$.}
\label{strength}
\end{figure}

Fig. \ref{strength} illustrates transformation of the mobility
curve with the increase of the strength of interaction. Again,
effective interaction strength $U_C$ affects mobility in the same
way as temperature. It is very interesting to compare three curves
in this figure: curves for $U_C=0$, $U_C=5$, and for a conceivable
time-of-flight (TOF) experiment, i.e. for $p\rightarrow 0$. The
first curve exactly corresponds to the results of Refs.
\cite{density1,density2}, where only filling of deep states by
non-interacting carriers has been taken into account, while the
second one is much more close to the real situation. Evidently,
when we consider the effect of carrier density on the mobility,
then the effect of charge-charge interactions is no less important
than the trivial filling of deep states.

\section{How to compare with experiment?}

One can suggest that carrier transport in organic field-effect
transistors (OFETs) should be a natural choice for comparison of
the simulation results with experiment \cite{FET1,FET2}.
Estimation of the carrier density in OFETs show that the density
as high as $3\times 10^{19}$ cm$^{-3}$ could be achieved
\cite{FET-density}, that for $a\approx 1$ nm corresponds to
$p\approx 0.03$. Experimental data for the particular OFET should
be compared with the TOF data for a sandwich device having
transport layer of the same material; in this way we could measure
 transport characteristics
(e.g., $\sigma$), relevant for the intrinsic disorder in the
material. Quite frequently, OFETs demonstrate mobilities much
higher that the mobilities measured in TOF experiments, and
usually mobility increases with the increase of $p$
\cite{FET-density}. This fact is in general agreement with the
model studied in this paper.

However, careful analysis reveals much more complicated situation.
Indeed, in many aspects OFETs are very far away from the model,
considered in the current study. First of all, in OFETs transport
occurs in a thin layer, close to the gate insulator. Quite
probably, especially in polymer devices, structure of this layer
differs from the structure of the same material in the bulk
(polymer chains could be arranged in a special way at the gate
insulator surface). This arrangement could provide more ordered
structure with less degree of energetic disorder, thus mobility
should be enhanced near the interface, but
 accumulation of surface defects and
impurities at the interface could lead to the decrease of the
mobility. Next, there is a clear indication that the roughness of
the organic semiconductor/dielectric interface affects carrier
mobility \cite{roughness}. At last, the very nature of a gate
dielectric (specifically, its polarity) affects carrier mobility
in OFETs, because a random orientation of polar groups in the
vicinity of a transport layer induces an additional energetic
disorder in semiconductor \cite{FET3,FET4}. In short, there are a
lot of reasons to believe that transport properties of OFETs are
too complicated to be directly compared with the results of this
study. We can only state that a significant increase of the
carrier mobility with the increase of carrier density in carefully
manufactured OFETs does not contradict the results of our study.

\section{Conclusion}
In conclusion, it was found that charge-charge interaction
significantly affects the carrier drift mobility and this effect
could not be described by the mobility dependence on average
electric field in the model of non-interacting carriers. Effect of
the charge-charge interaction is no less important for the true
description of the carrier mobility dependence on carrier density
than the previously considered effect of simple filling of deep
states.

This work was supported by the ISTC grant 3718 and RFBR grants
05-03-33206, 07-02-08484, and 05-03-90579-NNS.  This work made use
of the computation cluster of the Cornell Center for Materials
Research (CCMR) with support from the National Science Foundation
Materials Research Science and Engineering Centers (MRSEC) program
(DMR-0079992).

\end{document}